# Young stellar structures in four nearby galaxies

Petros Drazinos[1], A. Karampelas[1], E. Kontizas[2], M. Kontizas[1], A. Dapergolas[2],

E. Livanou[1], I. Bellas-Velidis[2]

1 - Department of Astrophysics, Astronomy & Mechanics, Faculty of Physics, University of Athens, GR-15783 Athens, Greece

2 - Institute for Astronomy and Astrophysics, National Observatory of Athens, P.O. Box 20048, GR-11810 Athens, Greece

## Abstract

A cluster finding method was developed and applied in four Local Group Galaxies (SMC, M31, M33 and NGC 6822). The aim is to study the young stellar population of these galaxies by identifying stellar structures in small and large scales. Also our aim is to assess the potential of using the observations of ESA's space mission Gaia for the study of nearby galaxies resolved in stars. The detection method used is a Hierarchical technique based on a modified friends of friends algorithm. The identified clusters are classified in five distinct categories according to their size. The data for our study were used from two ground based surveys, the Local Group Galaxy Survey and the Maggelanic Clouds Spectroscopic Survey. Relatively young main sequence stars were selected from the stellar catalogs and were used by the detection algorithm. Multiple young stellar structures were identified in all galaxies with size varying from very small scales of a few pc up to scales larger than 1 kpc. The same cluster finding method was used in six spiral galaxies observed with the Hubble Space Telescope in a previous study. The average size in each category of the identified structures in the Local Group galaxies presents values consistent with the identified structures in the relatively distant spiral galaxies. Most of the structures consist of stars within the observational limits of Gaia's instruments. It is expected that Gaia's observations will contribute significantly on the study of the young stellar population of nearby galaxies.

## 1. Introduction

The study of star forming regions is important as it can provide a better understanding on the stellar formation process and the structure of galaxies. Star forming regions are hierarchically structured on a large range of scales from superclouds down to associations associations (Elmegreen & Efremov 1996; Elmegreen 2006; Bastian et al. 2007; Gusev 2014) . The size of a structure corresponds to a timescale of star formation (Efremov & Elmegreen 1998). The large scale structures contain many sub structures with varying size down to the smallest scale, the associations which may be the building blocks of the structure of spiral galaxies Efremov & Chernin (1994). That makes the star forming regions and their stellar content an essential part in the process of galaxy evolution.

Different authors use the name "stellar association" for different kinds of stellar groups , Ivanov (1996), with dimensions depending on the identification criteria, distance of the galaxy, and image scale (Hodge 1986). The first who proposed the name stellar association was Abartsumian (1949) and later Blaauw (1964) described associations as a large gravitationally unbound stellar group of O and B stars. The processes used so far to identify an OB association were based on the subjective selection criteria of each observer (Lucke & Hodge 1970; Hodge 1986). As a consequence it was difficult to compare the results of different studies.

Many methods have been used for the study of stellar associations as it shown in Table 1. From detection by eye on photographic plates, star counts on photographic observations and isopleth contour maps (Maragoudaki et al 1998; Livanou et al 2006) to automated cluster finding techniques as the friends of friends (fof) algorithm on CCD observations, the Path Linkage Criterion (PLC) introduced by Battinelli (1991) and dentrograms, indicative Bastian et al. (2007); Maschberger et al (2010). In most of the cases as shown in Table 1 the mean size of the detected structures is less than 100 pc. Only in four studies the mean values are larger than 200 pc. Our aim is to identify structures in both small and large scales and in order to accomplish our goal we used a hierarchical method based on a modified fof algorithm Drazinos et al. (2013), hereafter Paper I. The fof algorithm was originally used to study galaxy clusters (Huchra & Geller 1982; Press & Davis 1982; Einasto et al. 1984). Our detection method was applied in six spiral galaxies observed by the Hubble Space Telescope (HST) Paper I detecting structures from ~ 30 pc up to a few kpc. In order to identify young stellar structures in a broad scale of sizes the algorithm automatically selects more than one search radii. The implementation of our method is further described in Section 3.

Clustering is an unsupervised process used to identify patterns in data using a similarity measure e.g. the euclidian distance between objects (Soni Madhulatha T. 2012; Yadav, Pandey & Mohanty 2015). It is widely used in many disciplines, biology, astronomy, statistics etc. Clustering techniques can be broadly divided into two major categories, Hierarchical and Partitional ( Agarwal, Afshar & Biswas 2011). Partitional techniques perform a single division of the data into k clusters which do not share common members. On the other hand the Hierarchical techniques perform a division of the data into clusters and subclusters and the process is performed in multiple steps. The two main categories of Hierarchical algorithms are the agglomerative and the divisive methods. The agglomerative methods start from the bottom, considering each member of the data set as a single cluster. In each step the singletons merge into larger clusters according to the similarity measure until the end of the process when all of the objects belong to a single cluster. The divisive methods start from the top, the whole data set is a single cluster. The single cluster is divided into smaller clusters in multiple steps until we end up with singletons (Kaufman & Rousseeuw 2005; Everitt et al 2011; Müllner D. 2011; Agarwal et al. 2011). The seven most common agglomerative methods depending on the method used for the similarity measure are termed single, complete, average (UPGMA), weighted (WPGMA, McQuitty),Ward, centroid (UPGMC) and median (WPGMC) linkage (Müllner D. 2011; Everitt et al 2011, Table 4.1). The fof algorithm is a Hierarchical agglomerative algorithm. The similarity measure in our method is the maximum great circle distance. Therefore two objects will be considered as members of the same group if their distance is less than a preselected value. Our method is a complete linkage algorithm which tends to find more symmetrical and compact clusters (Everitt et al 2011; Feigelson & Babu 2012).

Gaia, ESA's cornerstone mission will create a 3D map of the Milky Way with astrometric accuracies of about 10 μas at V = 15, observing about a billion stars in our Galaxy but also millions of extragalactic objects. Essentially Gaia will observe all objects in the sky up to a magnitude limit, G = 20 mag which transforms to Vlim = 20 − 25 mag depending on the color of the object. The main objective of Gaia is to provide data on the astrophysical characteristics and kinematics of the observed objects in order to study the dynamics and structure of our Galaxy (Perryman et al 2001; Prusti 2012; C.A.L. Bailer-Jones et al 2013). One of our goals is to assess whether the observations of Gaia can be used to study the stellar population of nearby galaxies. It is expected that Gaia will observe about nine million stars in the Magellanic Clouds Robin et al. (2012). Gaia uses two identical telescopes that are integrated with three distinct instruments, astrometric, photometric and spectroscopic that will provide the astrophysical and kinematical information of the detected objects. Positions, parallaxes and proper motions for all the observed objects will be provided by the astrometric instrument while the photometric instrument will provide low resolution spectra, one for the blue wavelengths (BP) covering the 330-680 nm range and one for the red wavelengths (RP) covering the 640-1000 nm range (Jordi et al 2010; de Bruijne 2012). From these low resolution spectra the spectral energy distribution of every detected object will be provided allowing for the classification of the object along with the determination of parameters as the effective temperature, metallicity, surface gravity, interstellar density for stars and monitoring for variability.

## 2. DATA

The study focuses on four nearby galaxies SMC, M31, M33 and NGC 6822. Observations from two ground based surveys were used for our method, the Magellanic Clouds Photometric Survey (MCPS), Zaritsky (2002), and the Local Group Galaxy Survey, Massey et al. (2006), available at ftp://ftp.lowell.edu/pub/massey/lgsurvey. These surveys were selected mainly because they cover most of the observed galaxies. In Paper I the observational data covered a small part of each studied galaxy. With these surveys we have the opportunity to use the detection algorithm in large observational areas and assess whether that affects the shape or the size of the detected structures. Also the wide coverage of both MCPS and LGGS provided a uniform large area coverage of the observed galaxies similar to Gaia's expected observations.

Archived data from the MCPS (Zaritsky 2002; Belcheva et al. 2011, private communication), were used for our study of the SMC. The MCPS data provided a number of advantages compared to similar surveys at the time as they are deeper or include a larger number of filters or cover a wider area of observation. The MCPS catalog covers an area of 4.5o $\times$ 4o for SMC. The observations of the SMC were made with the Las Campanas Swope 1m Telescope from 1996 November to 1999 December. Images were obtained in the Johnson's U, B, and V and the Gunn I and the incompleteness becomes significant at magnitudes fainter than V < 20. The scale of the Las Campanas is 0."7 / pixel, the typical seeing was ~ 1."5 and no scans with seeing worse than ~ 2."5 were accepted.

The data for M31, M33 and NGC 6822 were taken from the LGGS which obtained uniform large area coverage of star forming regions. M31 was imaged in ten fields covering 2.2 deg2, M33 was imaged in three fields covering 0.8 deg2 and NGC 6822 in a single 35' $\times$ 35' field. M31 and M33 were observed with the Kitt Peak National Observatory (KPNO) 4m telescope (scale : 0."261 /pixel) with good seeing conditions (< 1."0 − 1."2). NGC 6822, was observed with the Cerro Tololo Inter American Observatory (CTIO) 4m telescope (scale : 0."27/pixel).

## 3. METHOD

The cluster finding method used for this paper is a Hierarchical agglomerative complete linkage algorithm. The clustering in hierarchical agglomerative algorithms is a multi-step process. The algorithm divides a given data set into singletons where each object is considered a cluster. In our method singletons were not considered as valid clusters as we set a constrain for a cluster to have at least three members (n $\geq$ 3). Subsequently as the similarity measure (distance) changes the algorithm provides a new clustering of the data. The process continues until the selected distance is large enough that all data belong to a single cluster.

Our method is based on a fof algorithm which searches for pairs of objects that they are closer than a cut off distance limit. The main intention of the method is to apply an objective criterion to the stellar structure identification process. Two stars belong to the same group only if they lie at a distance less or equal to a predefined value of search radius. One important question needed to be addressed in this method is the value of search radius for which our star catalogue will be investigated.

Battinelli (1991) proposed to use the distance value that produces the maximum number of clusters (Figure 1). If only one distance value is selected the algorithm detects clusters of a small range of sizes. For M33 the maximum number of total groups, 467, is given for ds = 41 pc. Most of the detected structures for that specific ds value are up to 100 pc in size, 86.51%, of the total detected structures. Our aim is to detect a wide range of sizes. For M33 the maximum number of total groups, 467, is given for ds = 41 pc. Most of the detected structures for that specific ds value are up to 100 pc in size, 86.51% of the total detected structures. Our aim is to detect a wide range of structures from small to large scales. In order to achieve our goal we divide stellar structures in five categories. Based on the classification suggested by Efremov, Ivanov & Nikolov (1987) and on empirical evidence, Maragoudaki et al (1998); Maragoudaki et al (2001); Livanou et al (2006, 2007) and Karampelas et al (2009) the detected structures were divided into five categories depending on their size. Clusters (up to 30 pc), associations (30-100 pc), aggregates (100-300 pc), complexes (300-1000 pc) and supercomplexes (larger than 1000 pc). The algorithm at first identifies the distance value that produces the maximum number of structures in each of the five categories. For the identification process the method uses a discreet set of values, 1-200 pc with a step of 1 pc as shown in Figure 2. In each iteration of distance values the detected structures are classified into the categories mentioned above. When the five distinct values are identified the algorithm proceeds to the next step.

Clustering is performed for each of the five identified distance values. The final catalog of structures consists of all the groups identified in the process. It is possible that a group is detected in different distance values. In that case the multiple entries are removed and only the first identification of a group is retained to the final catalog. It should be noted that the centre of each group is defined as the mean values of Right Ascension (RA) and Declination (Dec) of its members. The size of each group is defined as the maximum distance found among the group's members. The results of the process and the plots used to select the five values are given in Figure 3.

## 3.1 Detection of young stellar structures

For SMC, M31, M33 and NGC 6822, from the data presented in section 2 we selected a sample of young main sequence bright stars in order to identify young stellar formations. For the SMC we selected stars with V<15.5, V-I<0.1 with ages estimated with theoretical isochrones (Marigo et al 2008) up to 100 Myr in order to have a sufficiently large sample for our algorithm and also to include larger stellar structures like supercomplexes (Figure 4, left). The structures identified in SMC consist mostly of stars within Gaia's RVS spectroscopic instrument detection limit, Grvs<17, (Jordi et al 2010) which will provide radial velocities allowing us the opportunity to evaluate the validity of our findings (whether stars are true members of the group they were assigned to or not), eliminating false detections from our identified groups. For the selection of relatively young main sequence stars in M31 and M33 the following constraints were used, V<20 and B-V<0.25 where from theoretical isochrones (Marigo et al 2008) the age of the sample was estimated to be 50 Myr at most (Figure 5). For NGC 6822 the constraints were slightly modified in order to increase the sample of stars, V<20, B-V<0.5 and from theoretical isochrones (Marigo et al 2008) the age was estimated to be more than 100 Myr (Figure 4, right).

## 3.2 Results

The identified groups vary in size, from a few pc up to a few kpc. In Figures 6, 7, 8 and 9 galaxies M31, M33, NGC 6822 and SMC, respectively, are mapped with the identified structures overplotted. In Table 2 we present for each galaxy, the search radii values for each category and in Table 3 a sample of the obtained catalog of the identified structures in SMC is given. Similar catalogs have been constructed for the other three galaxies and will be available online. The total number of identified groups for each size category (section 3) and the mean values of group size and number of members are given in Table 4.

## 3.3 Surface density

The surface density of the detected groups was estimated using a radius half the size of each group and assuming for simplicity that each group is a circular distribution occupying an area with surface equal to $\pi * radius^2$. The surface density for each galaxy is given in Figure 10. The surface density value seems to be correlated with the size of the group, small groups like clusters and associations have high surface density compared to the larger groups like supercomplexes. Similar to Bianchi et al. I (2012), Figure 11, where the surface density is higher for smaller groups. In Bianchi et al. I (2012) the observations were made with the HST which has a better resolution (0."05pixel−1) than KPNO (M31 and M33), CTIO (NGC 6822) and Las Campanas (SMC) providing a deeper coverage resulting in a larger number of observed stars. The link distances (equivalent to search radii or distance values) used in Bianchi et al were up to 22.8 pc, detecting probably what we define as associations or aggregates. Since the number of stars per detected group was much higher than in our study the surface density values were higher for smaller groups than the values presented in Figure 10.

# 4. DISCUSSION

The sizes of stellar associations in various galaxies from a number of publications are given in Gouliermis et al. (2003) (1). For M31, the sizes reported have a mean range of 80 to 83 pc, larger than the mean of 64 pc found in this study. The larger mean value can be explained from the difference in the identification technique (detection by eye) as in the case of Efremov et al. (1987). In M33 the variation in the mean value is wider, from 40 to 200 pc compared to 61 pc found with our technique. This range covers our size definitions of three of our categories, clusters, associations and aggregates. The lack of common definition of the identified structures can lead to a wide range of observed group sizes which can explain the difference in the reported mean size values. The various mean sizes published for SMC have a range from 46 to 77 pc compared to 60 pc from our identified groups. For NGC 6822 from Gouliermis et al. (2003) the range is 72 to 163 pc. Also Gouliermis et al. (2010) reported a mean size of 68 pc very close to our 67 pc average size, using a Nearest Neighbor method. It must be taken into account that in these publications, the definition of associations or other structures varied as well as the identification technique and the observational instrument. That can explain the wide range in some cases as in M33 and SMC in the reported mean sizes.

In Karampelas et al (2009) the detected structures in NGC 6822, SMC and LMC varied in two main size groupings. The first from ~ 150 to ~ 300-400 pc and the second up to ~ 800 pc. The range of these structures corresponds to the aggregates and complexes categories definition followed in this study. The size distribution of structures with size from 200 up to 800 pc for NGC 6822 and SMC is presented in Figure 11 and it is very similar to the one in Karampelas et al (2009) where there is a peak at ~ 200 pc about 34% of the detected groups in NGC 6822 and about 38% in SMC. Similarly in our study the size distribution for that size range peaks at ~ 200 pc, roughly the 32% and 47% of the groups respectively. It should be noted that in our study the N/Ntot values refer only to the 200-800 pc size range and not to the total number of detected groups, where in Karampelas et al (2009) the relative values refer to the total sample as there are only a few detections of smaller or larger groups, associations and supercomplexes. The size distribution of smaller structures is presented in Figure 12, for all four galaxies there is a peak below 30 pc (clusters) and around 40-80 pc (associations).

In Paper I were the data were obtained with HST, the targeted galaxies were at distances from 6.3 to 15.9 pc and were partially observed. The mean value of associations in each galaxy varied similarly to the four galaxies studied in this paper, from 62 to 75 pc (Paper I, Table 5). Comparing the range of mean values of structures larger than associations of the galaxies from Paper I with the relative values of the four nearby galaxies there are no significant differences except for supercomplexes. In the three of the four nearby galaxies the identified supercomplexes had mean values over 2 kpc where in the more distant galaxies the mean value varied from 1.4 to 1.8 kpc. A possible explanation for the difference in supercomplexes is the partial coverage of the distant galaxies. The size of the observed field was limited compared to the nearby galaxies where M33 and NGC 6822 where completely covered by the observations. The limited field can affect the algorithm by not detecting a sufficient number of large structures that can increase the mean value or by not detecting large structures at all. In order to eliminate the effect of extreme findings (very large structures which can increase the average value significantly) we plot the median values of all categories for each of the ten galaxies. The resulting plot is presented in Figure 14. The median sizes of supercomplexes are in the range ~ 1200-1700 pc for most of the nearby and distant galaxies. This is an indication that the wide field of the nearby galaxies allowed the detection of larger supercomplexes than in a partially observed field. The other categories are unaffected by the size of the observed field.

In Figure 13 the average size is given for each category for all galaxies studied with our method. In total, ten galaxies were studied in a variety of distances, using data from three surveys and four different instruments. The average size of associations has a range from 60 to 75 pc, of aggregates from 160 to 182 pc, of complexes from 481 to 537 and of supercomplexes from 1424 to 2408 pc. The detection of structures with diameters less than 30 pc was not attempted in the study of the six HST galaxies so they are not presented in this figure. It should be noted that the mean values for each category do not seem to depend on the distance of the galaxy, the observational instrument and the galaxy type although the sample is small (two Irregulars and eight Spirals).

Similar findings of an approximately constant mean value of structures with size less than 200 pc were reported in Bresolin, Kennicut & Stetson (1996) and in Bresolin et al (1998) for twelve galaxies including all of the ten galaxies studied in this paper and in Paper I. The mean values are higher than our estimations as in both cases all structures with diameters up to 200 pc were considered as associations which is essentially double in size than in our definition. Also in Ivanov (1996) the mean value of detected structures in eight galaxies (including M31, M33 and NGC 6822) ranged from 72 to 114 pc with a mean value of 84 pc. Structures detected in NGC 628 (Gusev 2014) Figure 6 and Table 2, present a size distribution similar to our findings. In the size distribution (Figure 6) there are three distinct peaks, below 100 pc (associations and clusters), around 250 pc (aggregates) and around 600 pc (complexes) as in Figures 11 and 12. The mean sizes of associations is given around 60-70 pc as was estimated in Figure 13 for the ten galaxies. The mean sizes for aggregates ~ 234pc and for complexes ~ 601pc are slightly larger than our estimations, 160-182 pc and 481-537 pc respectively, but well within the range of each size category.

Also in similar studies of nearby galaxies, Efremov et al. (1987); Magnier et al (1993); Battinelli (1991); Battinelli, Efremov & Magnier (1996);Gouliermis et al. (2003, 2010), and for galaxies at greater distances like in Paper I, Bresolin et al. (1996); Bresolin et al (1998); Gusev (2014), distinct scales of structures have been reported. Specifically for associations, aggregates and complexes which can be found in these studies under different definitions. The smallest and larger structures, like clusters and supercomplexes are not mentioned frequently. That can be explained because the clusters due to their small size are at the limit of an instruments resolution. The supercomplexes are at the other end of scale and from empirical evidence the detection of such structures can depend heavily on the used method where in most cases, medium sizes structures up to 300 pc can be easier to identify. In algorithms similar to ours usually one distance value is used for the clustering process. When the distance value producing the maximum number of groups is used in the algorithm the majority of the detected groups are associations, aggregates or complexes resulting in a mean value close to medium sized structures. ESA's space mission Gaia, described in detail in Perryman et al (2001) and Prusti (2012) will offer in the upcoming years the opportunity to study the star forming regions of nearby galaxies like the Magellanic Clouds. It is expected that about nine million stars will be observed in the LMC and SMC (Robin et al. 2012) and more than 20,000 stars in M31, M33 and NGC 6822 (Drazinos et al. 2014). Gaia will offer homogenous observational data in resolution compared to the HST. Spectroscopic observations for stars G > 17 mag will provide radial velocities which will allow us to add another criterion in the detection process of young stellar clusters. Using the radial velocities we should be able to eliminate false detections of structures and construct a more robust algorithm. Also with the astrophysical information (for example, effective temperature, surface gravity, metallicity) that will be provided for the observed stars can be used for a further analysis of the detected structures. It must be taken into account that most of the detected clusters in this paper consist of stars within Gaia's photometric instrument and especially in the case of SMC the structures fall within the detection limit of the high resolution spectroscopic instrument. The higher resolution of Gaia than the ground based instruments could significantly increase the number of extragalactic stellar observations than the estimated numbers given above.

## 5. CONCLUSIONS

Star formation regions in four nearby galaxies, SMC, NGC 6822, M31 and M33 were studied aiming for the detection of relatively young stellar structures in a broad scale of sizes. The stellar structures were detected using a Hierarchical cluster finding method based on a modified friends of friends algorithm. Our method was developed to address specifically the issue of detecting stellar structures in a large variety of sizes. Structures are divided in five distinct categories according to their size and the algorithm selects automatically one search radius for each category. The data for the study were taken from two ground based surveys LGGS and MCPS. In all four galaxies studied stellar structures were identified with sizes varying from a few pc up to a few kpc. The range of the mean size of the stellar structures found are consistent with previous findings in relatively distant spiral galaxies observed with HST (Paper I) using the same method. The mean size range of associations, aggregates, complexes is in agreement with the sizes found in literature although a variety of definitions were used to describe detected structures in different studies. It seems that young stellar structures can be divided into distinct scales as the mean size of each category did not differ significantly for each of the ten galaxies presented in Figures 13 and 14 despite the difference in galaxy type, the distance of the galaxy and instrument of observation. The smaller structures in the four nearby galaxies have higher surface density values than larger sized groups. Hierarchical structure was indicated in all four galaxies as was also the case in Paper I. Small and dense structures like clusters and associations seem to be engulfed by larger and loose structures like complexes and supercomplexes. The upcoming data from Gaia space mission are expected to contribute in the study of the stellar population of the Local Group galaxies. All the detected structures are within the detection limits of Gaia's instruments. The characteristics and kinematics of the observed stars will provide the means to improve the detection algorithm performance and for further analysis of the detected structures.

## ACKNOWLEDGEMENTS

The authors would like to thank PRODEX/ESA for financial support. Also the authors would like to thank the National and Kapodistrian University of Athens (NKUA) and the National Observatory of Athens (NOA) for their support.

**Table 1.** Detection methods and results (Gouliermis et al. 2003).

| Galaxy Name | Hubble Type | Number | min | Size (pc) mean | max | References | Detection Method |
|---|---|---|---|---|---|---|---|
| Sextans A | E | 3 | | 93 | | Ivanov (1996) | 6 |
| M31 | Sb | 210 | 20 | 80 | | Efremov et al. (1987) | 1 |
| | | 15 | | 83 | | Ivanov (1996) | 6 |
| NGC 7331 | Sb | 142 | | 440 | | Hodge (1986) | 1 |
| M33 | Sc | 143 | | 200 | | Humphreys & Sandage (1980) | 1 |
| | | 460 | 30 | 80 | 270 | Ivanov (1987) | 1 |
| | | 289 | 6 | 66 | 305 | Ivanov (1991) | 5 |
| | | 8 | | 87 | | Ivanov (1996) | 6 |
| | | 41 | 10 | 40 | 120 | Wilson (1991) | 3 |
| NGC 2403 | Sc | 88 | 160 | 348 | 600 | Hodge (1985a) | 1 |
| NGC 4303 | SBbc | 235 | | 290 | | Hodge (1986) | 1 |
| LMC | Irr | 122 | 15 | 78 | 150 | Lucke & Hodge (1970) | 1 |
| | | 2883 | 5 | 18 | 272 | Bica et al. (1999) | 1 |
| | | 153 | 21 | 86 | 190 | Gouliermis (2003) | 7 |
| | | 24 | | 250 | | Maragoudaki (2009) | 8 |
| | | 5 | | 164 | | Livanou (2006) | 8 |
| SMC | Irr | 70 | 18 | 77 | 180 | Hodge (1985b) | 1 |
| | | 31 | 50 | 90 | 270 | Battinelli (1991) | 4 |
| | | 200 | 9 | 46 | 234 | Bica & Schmitt (1995) | 1 |
| | | | | ~ 150 | | Livanou (2006) | 8 |
| NGC 6822 | Ir+ | 16 | 48 | 163 | 305 | Hodge (1977) | 2 |
| | | | | 150 | | Karampelas (2009) | 8 |
| | | 6 | | 72 | | Ivanov (1996) | 6 |
| IC 1613 | Irr | 20 | 68 | 164 | 485 | Hodge (1978) | 2 |
| | | 6 | | 83 | | Ivanov (1996) | 6 |
| Pegasus | Irr | 3 | | 65 | | Ivanov (1996) | 6 |
| GR8 | Irr | 3 | | 114 | | Ivanov (1996) | 6 |
| HoIX | Im | 3 | | 72 | | Ivanov (1996) | 6 |

DETECTION METHODS EXPLANATIONS
1: Detection by eye on photographic plates or films.
2: Detection by eye using star counts from photoelectric and photographic observations.
3: Friends of friends grouping algorithm on stars from CCD observations.
4: Path Linkage Criterion applied on O-B2 stars selected from objective-prism observations.
5: Cluster analysis technique on stars from photographic observations.
6: Automated cluster analysis technique on OB stars selected from CCD observations.
7: Objective statistical method based on star counts from photographic stellar catalogs.
8: Isopleth contour maps.

**Table 2.** Values of search radius (Distance) selected from the algorithm for each size category.

| Galaxy | Clusters (pc) | Associations (pc) | Aggregates (pc) | Complexes (pc) | Supercomplexes (pc) |
|---|---|---|---|---|---|
| SMC | 21 | 34 | 46 | 54 | 73 |
| M31 | 19 | 61 | 116 | 234 | 296 |
| M33 | 14 | 41 | 76 | 135 | 161 |
| NGC 6822 | 19 | 63 | 115 | 144 | 193 |

**Table 3.** Catalog sample of groups identified in SMC (Type of group : Clusters (0), Associations (1), Aggregates (2), Complexes (3), Supercomplexes (4)).

| Group Index | RA (deg) (J2000.0) | DEC (deg) (J2000.0) | Members | Size (pc) | Type |
|---|---|---|---|---|---|
| 1 | 9.302047 | -72.993367 | 3 | 35.591404 | 1 |
| 2 | 11.857450 | -73.107583 | 6 | 37.920717 | 1 |
| 3 | 11.871925 | -73.217400 | 4 | 27.360612 | 0 |
| 4 | 11.895733 | -73.146533 | 3 | 28.19969 | 0 |
| 5 | 11.943700 | -73.133700 | 3 | 25.173601 | 0 |
| 6 | 11.975867 | -73.103233 | 3 | 20.062199 | 0 |
| 7 | 11.987433 | -73.424400 | 3 | 23.918704 | 0 |
| 8 | 12.011467 | -73.406900 | 3 | 18.795968 | 0 |
| 9 | 12.156767 | -73.423067 | 3 | 10.772526 | 0 |
| 10 | 12.197640 | -73.427580 | 5 | 22.069062 | 0 |

**Table 4.** Number of groups for each group type and their characteristics for each galaxy.

| Galaxy | Type | Number of groups | Average size (pc) | Average number of members (pc) | Median size (pc) |
|---|---|---|---|---|---|
| SMC | | | | | |
| | Clusters | 116 | 22 | 3 | 24 |
| | Associations | 346 | 60 | 5 | 57 |
| | Aggregates | 151 | 160 | 13 | 148 |
| | Complexes | 27 | 481 | 59 | 448 |
| | Supercomplexes | 6 | 1958 | 518 | 1640 |
| M31 | | | | | |
| | Clusters | 154 | 13 | 4 | 11 |
| | Associations | 230 | 64 | 4 | 65 |
| | Aggregates | 354 | 182 | 6 | 174 |
| | Complexes | 228 | 537 | 13 | 485 |
| | Supercomplexes | 61 | 2257 | 110 | 1700 |
| M33 | | | | | |
| | Clusters | 354 | 15 | 4 | 14 |
| | Associations | 371 | 61 | 5 | 60 |
| | Aggregates | 268 | 168 | 10 | 154 |
| | Complexes | 105 | 482 | 33 | 438 |
| | Supercomplexes | 19 | 2080 | 553 | 1374 |
| NGC 6822 | | | | | |
| | Clusters | 47 | 16 | 4 | 17 |
| | Associations | 69 | 67 | 5 | 70 |
| | Aggregates | 70 | 165 | 5 | 150 |
| | Complexes | 28 | 522 | 17 | 517 |
| | Supercomplexes | 7 | 2408 | 520 | 2011 |

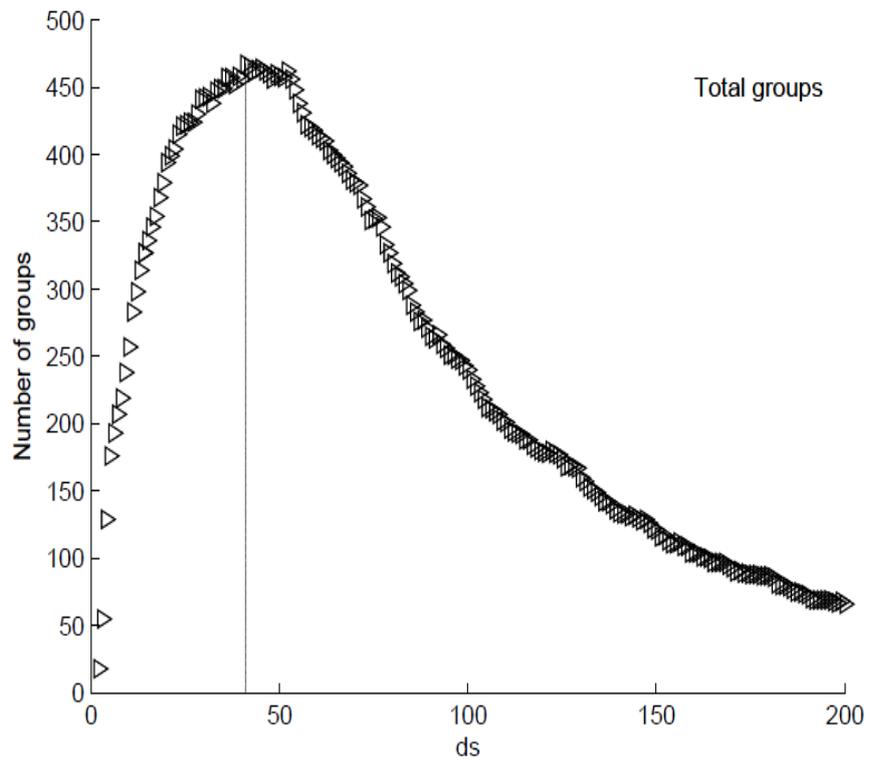

Figure 1. Plot of the distance value and the total number of structures detected by the algorithm in M33. The distance value that produces the maximum number of structures is selected.

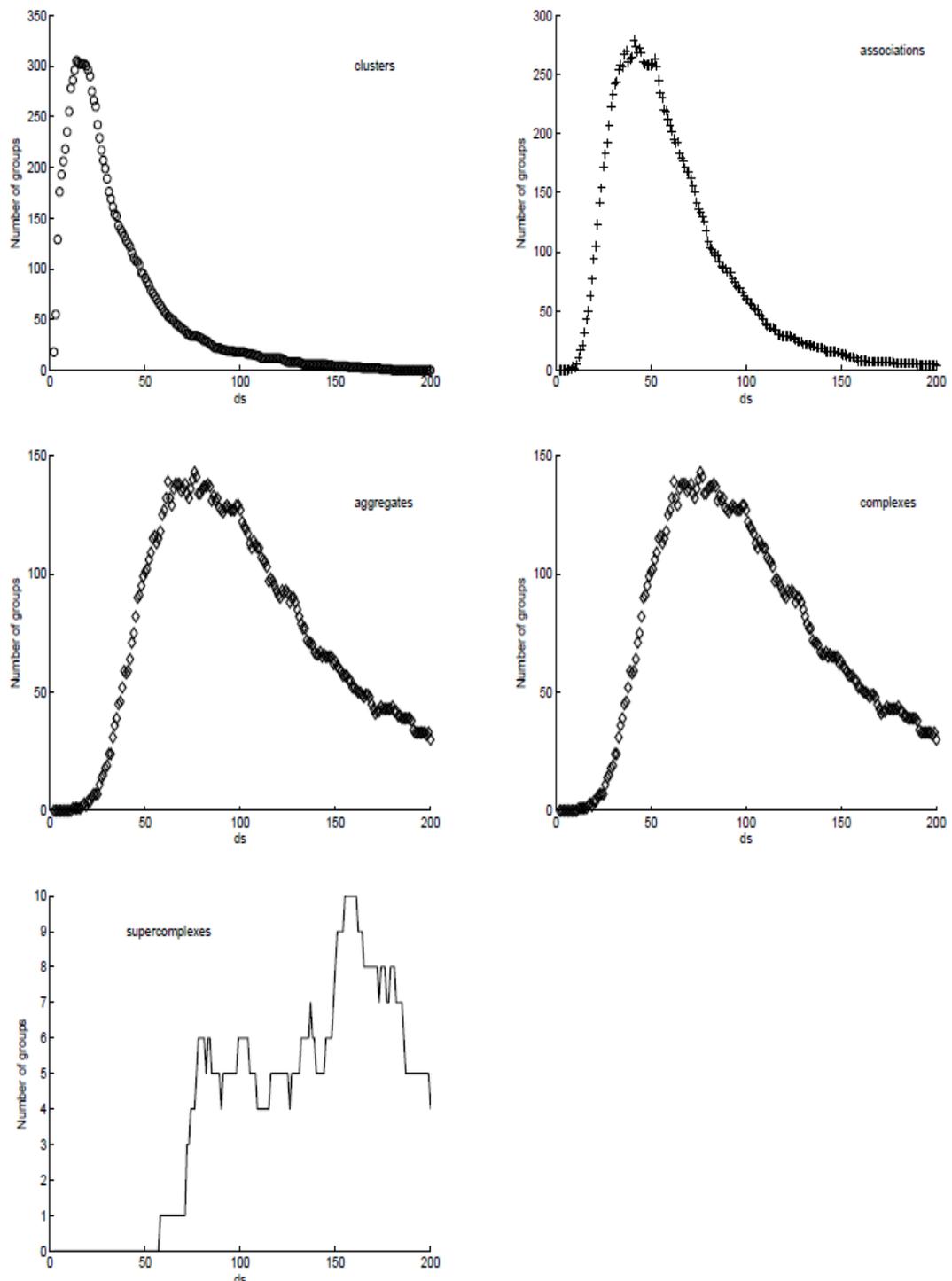

**Figure 2.** Selection of a single distance value for each of the size categories in M33.

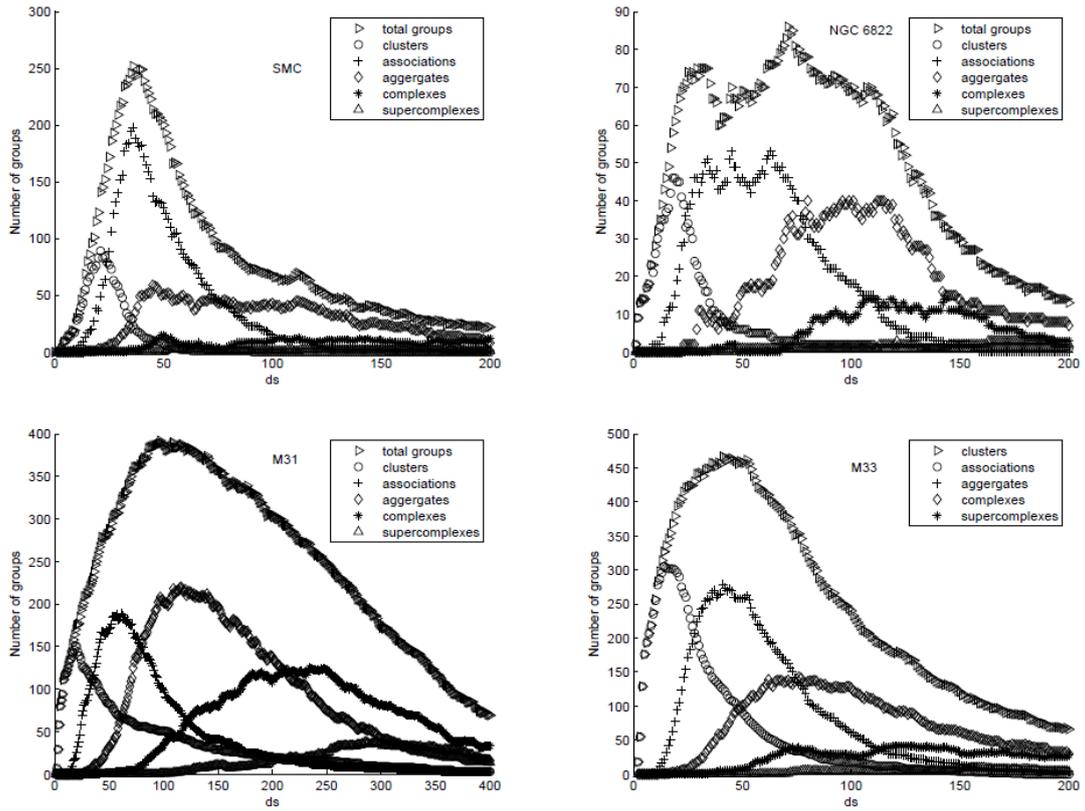

Figure 3. Number of groups identified for each search radius value for each of the studied galaxies.

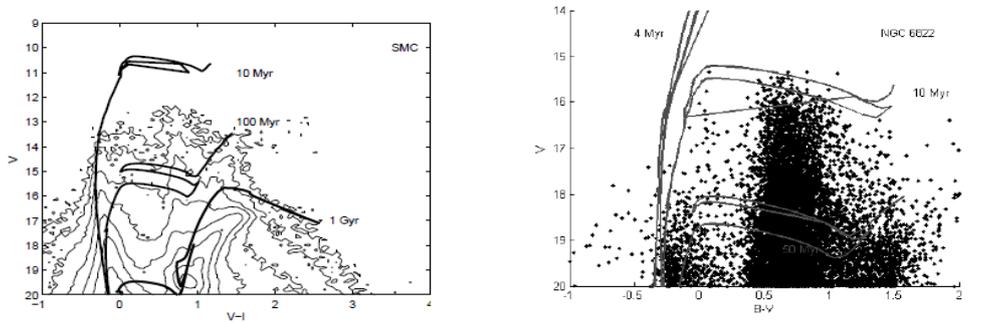

Figure 4. CMD's of irregulars SMC with isochrones for 10 Myr, 100 Myr and 1 Gyr (left) and NGC 6822 with isochrones for 4, 10 and 100 Myr (right).

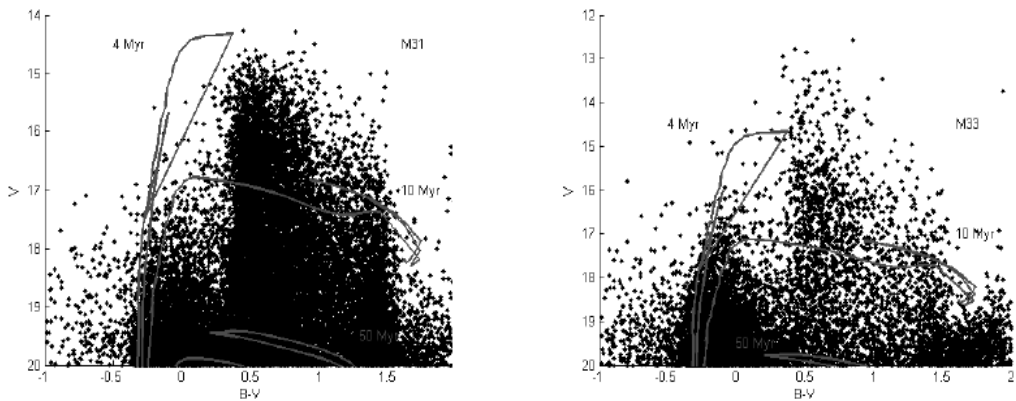

Figure 5. CMD's of spirals M31, M33 with isochrones for 4, 10 and 50 Myr and CMD of NGC 6822.

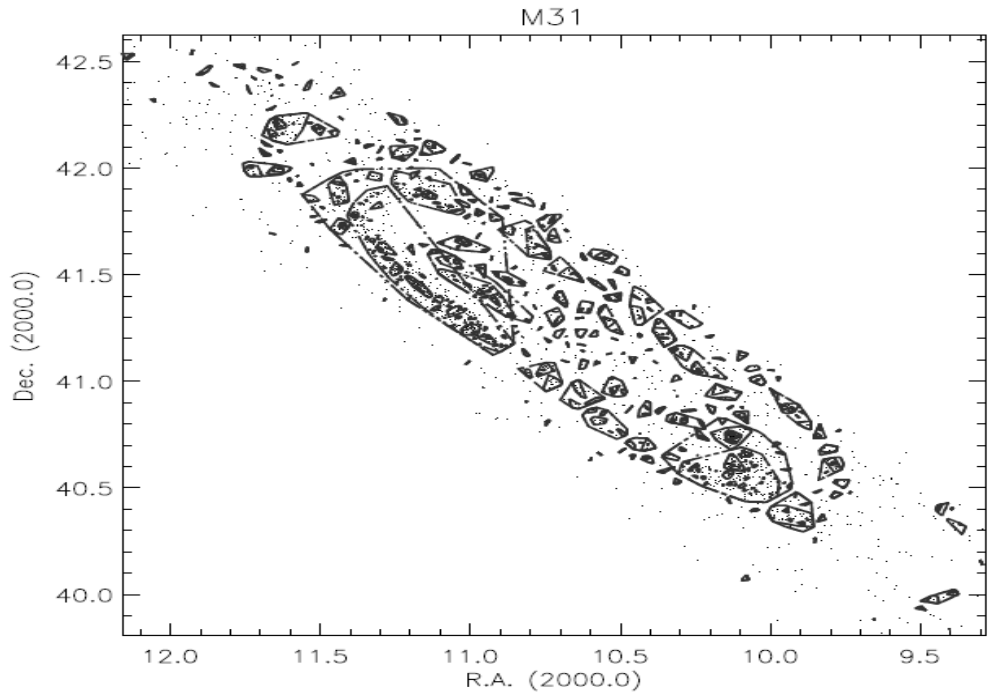

Figure 6. Young stellar structures in M31.

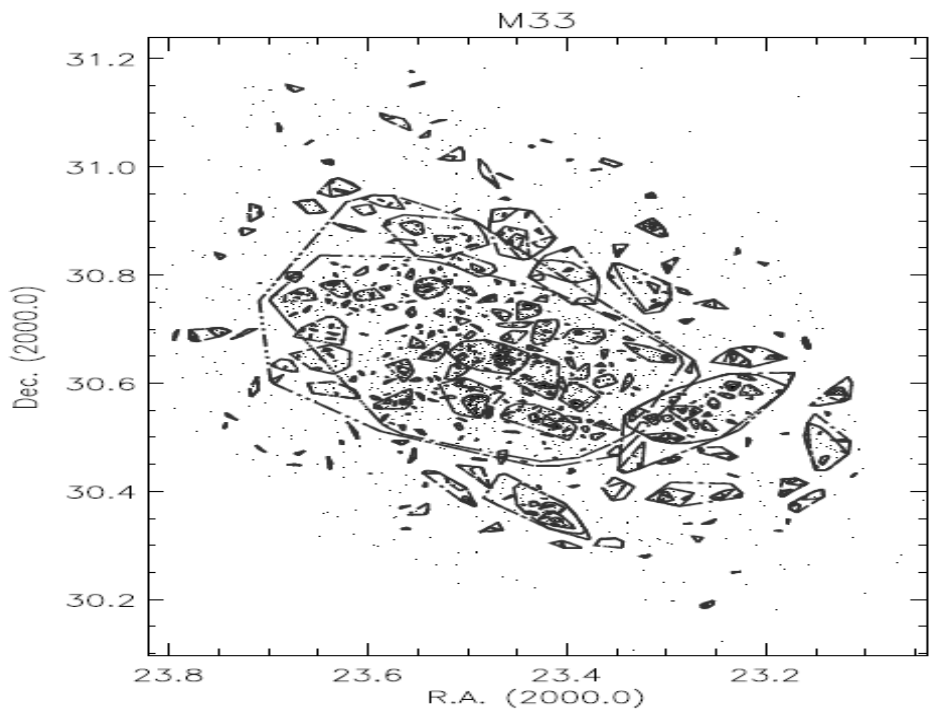

Figure 7. Young stellar structures in M33.

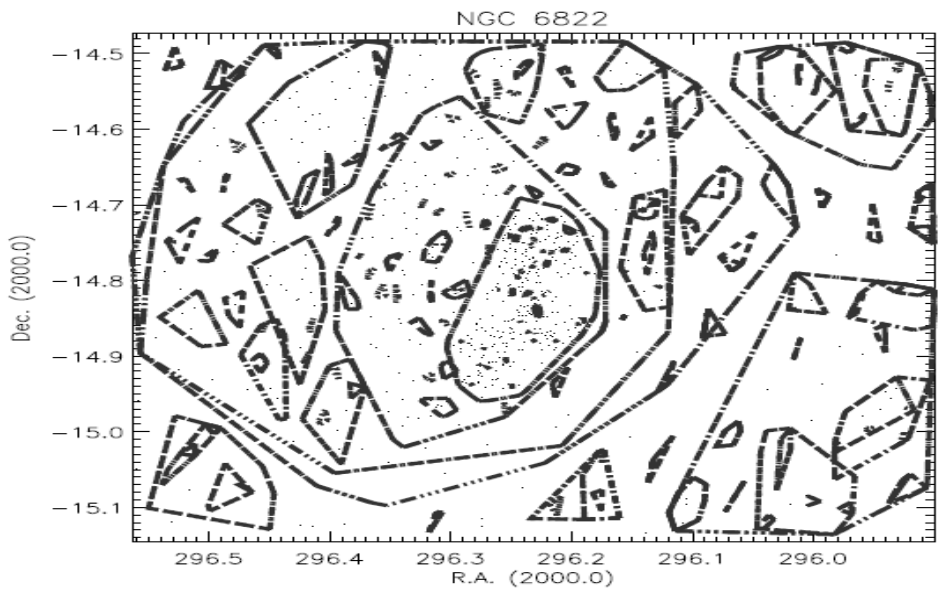

**Figure 8.** Young stellar structures in NGC 6822.

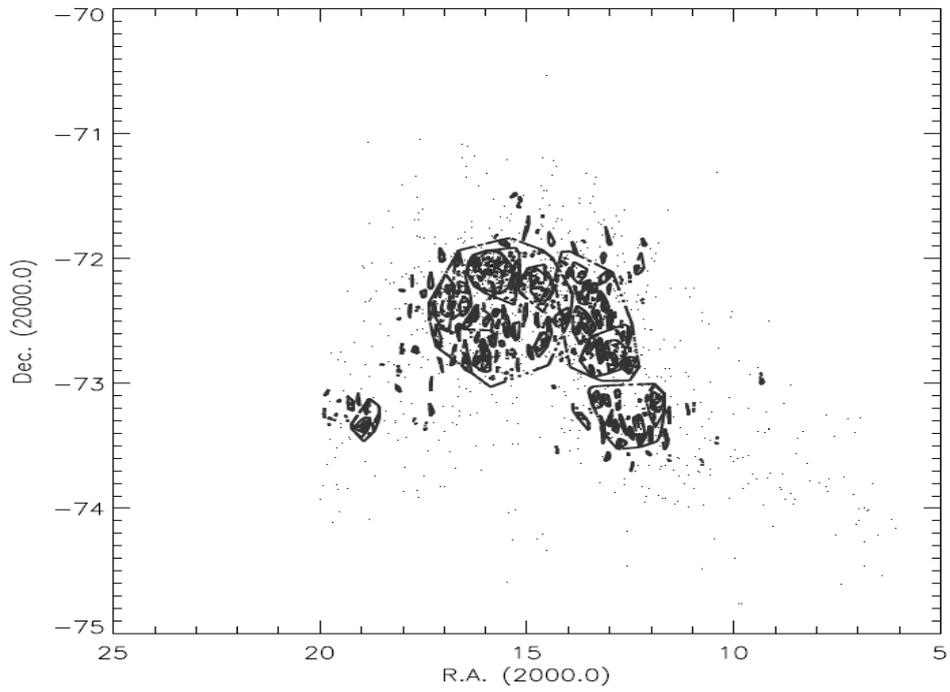

**Figure 9.** Young stellar structures in SMC.

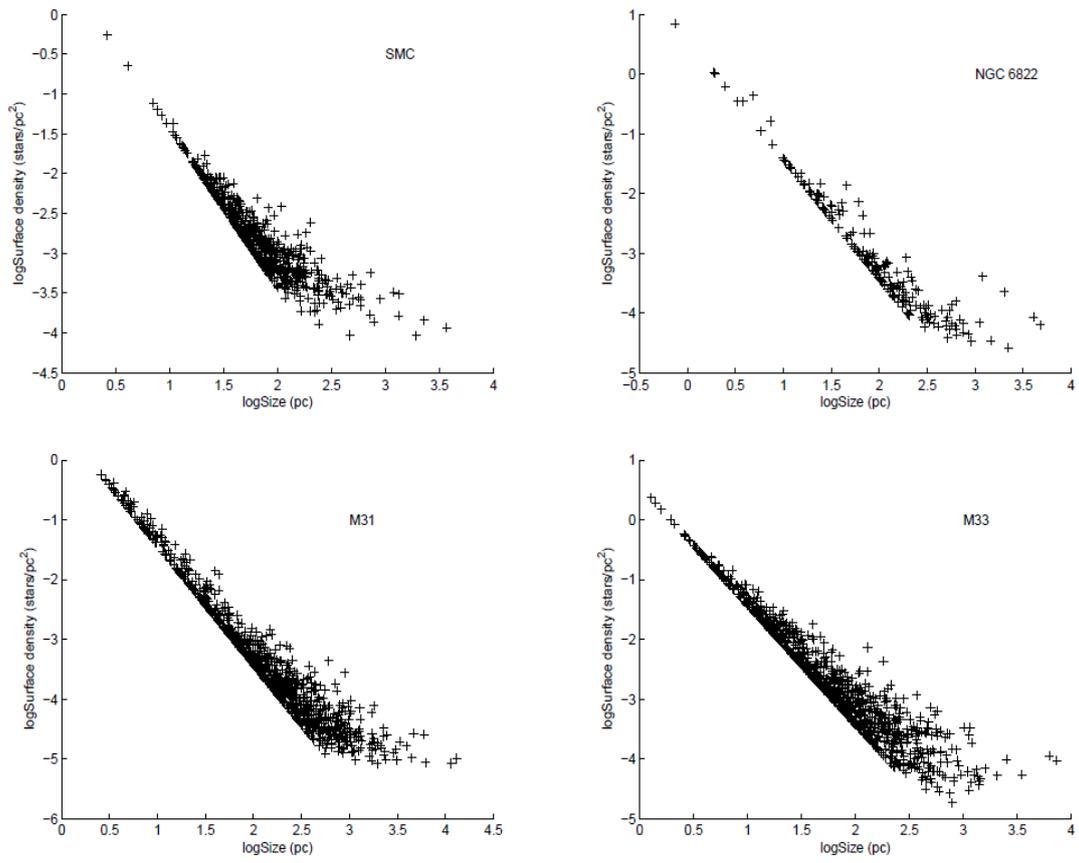

**Figure 10.** Surface density estimated for SMC, NGC 6822, M31 and M33.

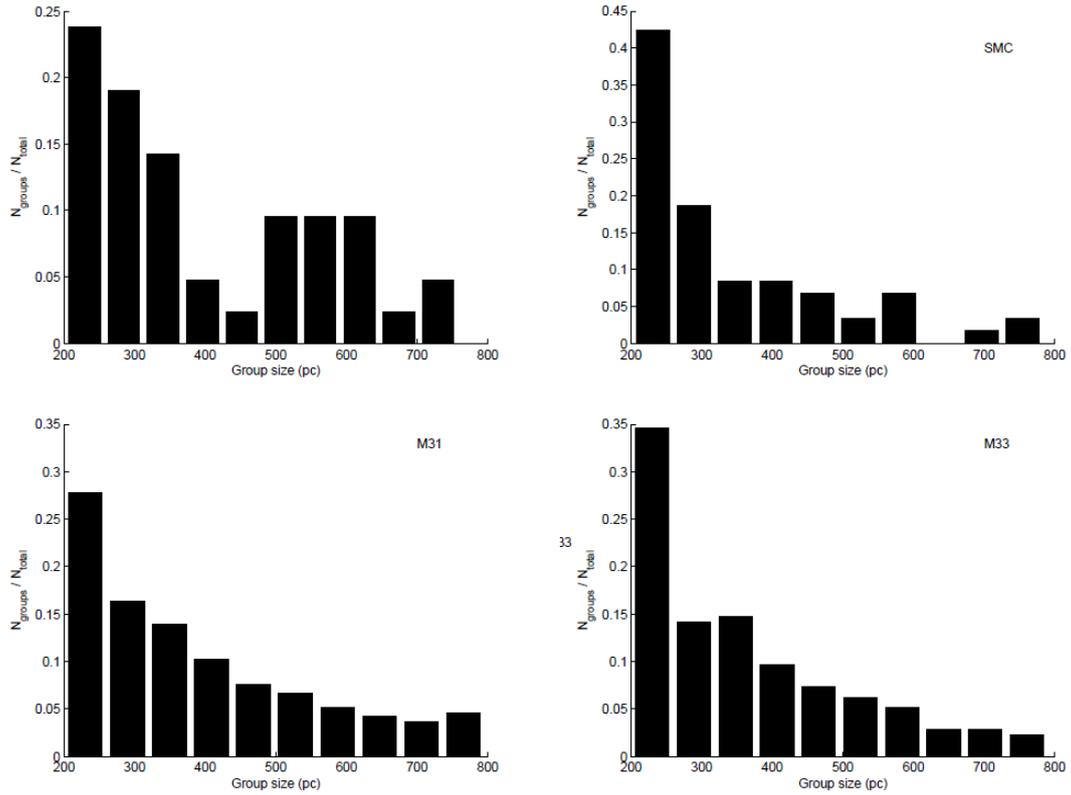

**Figure 11.** Size distribution for detected groups with size from 200 to 800 pc in NGC 6822, SMC, M31 and M33.

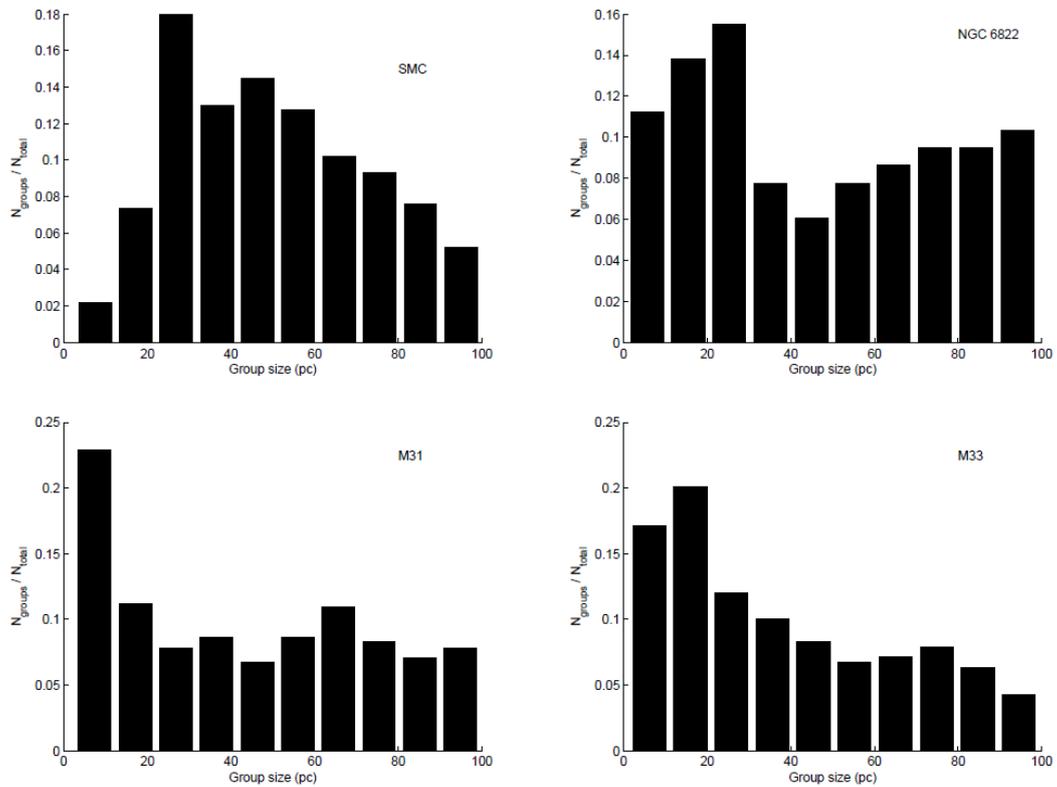

**Figure 12.** Size distribution of young stellar structures with size up to 100 pc found in SMC, NGC 6822, M31 and M33.

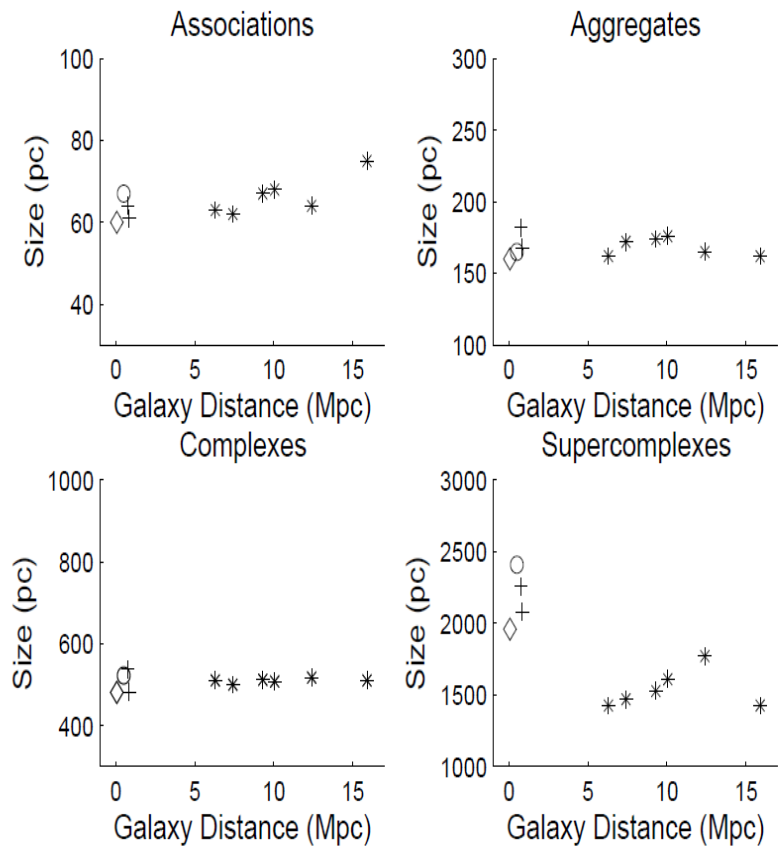

Figure 13. The mean size range of detected structures for ten galaxies in various distances observed with four different instruments. From top left and clockwise, mean values of associations, aggregates, supercomplexes and complexes. Eight of the studied galaxies (M31, M33 and six from Paper I) are spirals observed with HST (∗) and KPNO (+). Two (SMC and NGC 6822) are irregulars observed with CTIO (◦) and Las Campanas (⋄).

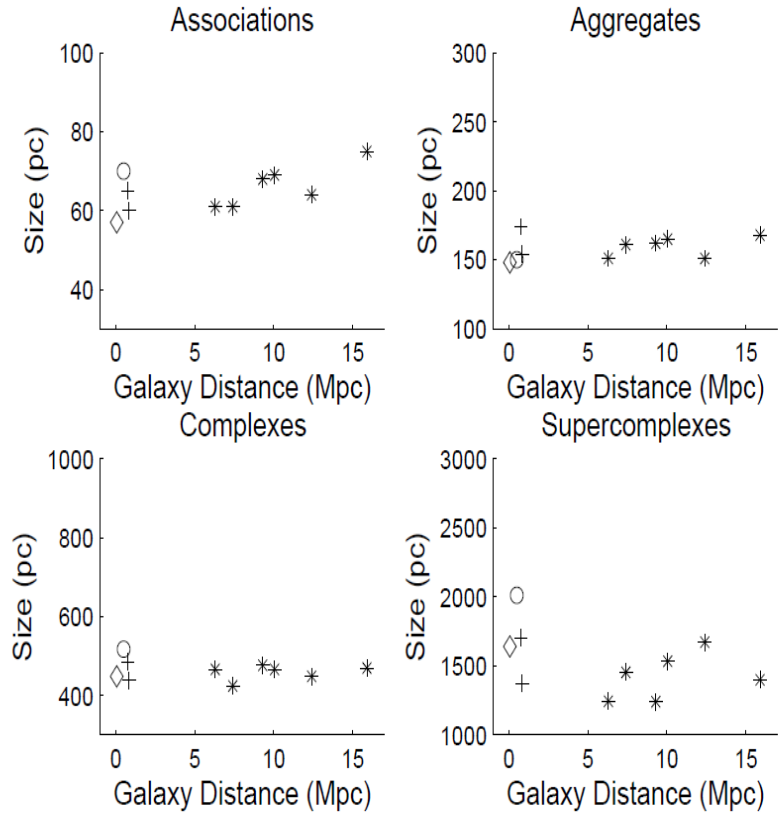

**Figure 14.** The median size range of detected structures for ten galaxies in various distances observed with four different instruments. From top left and clockwise, mean values of associations, aggregates, supercomplexes and complexes. Eight of the studied galaxies (M31, M33 and six from Paper I) are spirals observed with HST (∗) and KPNO (+). Two (SMC and NGC 6822) are irregulars observed with CTIO (○) and Las Campanas (◇).